\documentclass[12pt,english]{paper}
\usepackage[T1]{fontenc}
\usepackage[latin1]{inputenc}
\usepackage{graphicx}

\makeatletter

\usepackage{lineno}

\makeatletter

\makeatletter

\makeatletter

\usepackage{geometry}

\makeatletter

\usepackage{epsfig}

\makeatother

\makeatother

\makeatother

\makeatother

\usepackage{babel}
\makeatother
\begin{document}

\title{Applying Bayesian Neural Networks to Event Reconstruction in Reactor
Neutrino Experiments }

\author{Ye Xu%
\thanks{Corresponding author, e-mail address: xuye76@nankai.edu.cn%
}, Weiwei Xu, Yixiong Meng, Kaien Zhu, Wei Xu}

\maketitle
Department of Physics, Nankai University, Tianjin 300071, People's
Republic of China

\begin{abstract}
A toy detector has been designed to simulate central detectors in
reactor neutrino experiments in the paper. The electron samples from
the Monte-Carlo simulation of the toy detector have been
reconstructed by the method of Bayesian neural networks (BNN) and
the standard algorithm, a maximum likelihood method (MLD),
respectively. The result of the event reconstruction using BNN has
been compared with the one using MLD. Compared to MLD, the
uncertainties of the electron vertex are not improved, but the
energy resolutions are significantly improved using BNN. And the
improvement is more obvious for the high energy electrons than the
low energy ones.
\end{abstract}
\begin{flushleft}
\par\end{flushleft}

\section{Introduction}

The main goals of reactor neutrino experiments are to detect $\bar{\nu_{e}}\rightarrow\bar{\nu_{x}}$
oscillation and precisely measure the mixing angle of neutrino oscillation
$\theta_{13}$. The experiment is designed to detect reactor $\bar{\nu_{e}}$'s
via the inverse $\beta$-decay reaction

\begin{center}
$\bar{\nu_{e}}+p\rightarrow e^{+}+n$
\par\end{center}

\begin{flushleft}
The signature is a delayed coincidence between $e^{+}$ and the
neutron captured signals. It is very important to reconstruct the
energy and the vertex of a signal detected in the experiments. The
standard algorithm of the event reconstruction in the experiments is
a maximum likelihood method (MLD from now on). But the method of
Bayesian neural networks (BNN from now on)\cite{key-1} is more
suitable than MLD for the event reconstruction of reactor neutrino
experiments. BNN is an algorithm of the neural networks trained by
Bayesian statistics. It is not only a non-linear function, but also
controls model complexity. So its flexibility makes it possible to
discover more general relationships in data than the traditional
statistical methods and its preferring simple models make it
possible to solve the over-fitting problem better than the general
neural networks\cite{key-2}. In this paper, BNN is applied to the
event reconstruction of the electron samples from the Monte-Carlo
simulation of a toy detector of reactor neutrino experiments. And
the result of the event reconstruction using BNN is compared with
the one using MLD.
\par\end{flushleft}

\section{The Regression with Bayesian Neural Networks\cite{key-1,key-3}}

The idea of BNN is to regard the process of training a neural network
as a Bayesian inference. Bayes' theorem is used to assign a posterior
density to each point, $\bar{\theta}$, in the parameter space of
the neural networks. Each point $\bar{\theta}$ denotes a neural network.
In the method of BNN, one performs a weighted average over all points
in the parameter space of the neural network, that is, all neural
networks. The methods make use of training data \{($x_{1}$,$t_{1}$),
($x_{2}$,$t_{2}$),...,($x_{n}$,$t_{n}$)\}, where $t_{i}$ is the
known target value associated with data $x_{i}$, which has $P$ components
if there are $P$ input values in the regression. That is the set
of data $x=$($x_{1}$,$x_{2}$,...,$x_{n}$) which corresponds to
the set of target $t=$($t_{1}$,$t_{2}$,...,$t_{n}$). The posterior
density assigned to the point $\bar{\theta}$, that is, to a neural
network, is given by Bayes' theorem

\begin{center}
\begin{equation}
p\left(\bar{\theta}\mid x,t\right)=\frac{\mathit{p\left(x,t\mid\bar{\theta}\right)p\left(\bar{\theta}\right)}}{p\left(x,t\right)}=\frac{p\left(t\mid x,\bar{\theta}\right)p\left(x\mid\bar{\theta}\right)p\left(\bar{\theta}\right)}{p\left(t\mid x\right)p\left(x\right)}=\frac{\mathit{p\left(t\mid x,\bar{\theta}\right)p\left(\bar{\theta}\right)}}{p\left(t\mid x\right)}\end{equation}

\par\end{center}

\begin{flushleft}
where data $x$ do not depend on $\bar{\theta}$, so $p\left(x\mid\theta\right)=p\left(x\right)$.
We need the likelihood $p\left(t\mid x,\bar{\theta}\right)$ and the
prior density $p\left(\bar{\theta}\right)$, in order to assign the
posterior density $p\left(\bar{\theta}\mid x,t\right)$to a neural
network defined by the point $\bar{\theta}$. $p\left(t\mid x\right)$
is called evidence and plays the role of a normalizing constant, so
we ignore the evidence. That is,
\par\end{flushleft}

\begin{center}
\begin{equation}
Posterior\propto Likelihood\times Prior\end{equation}

\par\end{center}

\begin{flushleft}
We consider a class of neural networks defined by the function
\par\end{flushleft}

\begin{center}
\begin{equation}
y\left(x,\bar{\theta}\right)=b+{\textstyle {\displaystyle \sum_{j=1}^{H}v_{j}sin\left(a_{j}+\sum_{i=1}^{P}u_{ij}x_{i}\right)}}\end{equation}
 .
\par\end{center}

\begin{flushleft}
The neural networks have $P$ inputs, a single hidden layer of $H$
hidden nodes and one output. In the particular BNN described here,
each neural network has the same structure. The parameter $u_{ij}$
and $v_{j}$ are called the weights and $a_{j}$ and $b$ are called
the biases. Both sets of parameters are generally referred to collectively
as the weights of the BNN, $\bar{\theta}$. $y\left(x,\bar{\theta}\right)$
is the predicted target value. We assume that the noise on target
values can be modeled by the Gaussian distribution. So the likelihood
of $n$ training events is
\par\end{flushleft}

\begin{center}
\begin{equation}
p\left(t\mid x,\bar{\theta}\right)=\prod_{i=1}^{n}exp[-((t_{i}-y\left(x_{i},\bar{\theta}\right))^{2}/2\sigma^{2}]=exp[-\sum_{i=1}^{n}(t_{i}-y\left(x_{i},\bar{\theta}\right)/2\sigma^{2})]\end{equation}

\par\end{center}

\begin{flushleft}
where $t_{i}$ is the target value, and $\sigma$ is the standard
deviation of the noise. It has been assumed that the events are independent
with each other. Then, the likelihood of the predicted target value
is computed by Eq. (4).
\par\end{flushleft}

\begin{flushleft}
We get the likelihood, meanwhile we need the prior to compute the
posterior density. But the choice of prior is not obvious. However,
experience suggests a reasonable class is the priors of Gaussian class
centered at zero, which prefers smaller rather than larger weights,
because smaller weights yield smoother fits to data . In the paper,
a Gaussian prior is specified for each weight using the Bayesian neural
networks package of Radford Neal%
\footnote{R. M. Neal, \emph{Software for Flexible Bayesian Modeling and Markov
Chain Sampling}, http://www.cs.utoronto.ca/\textasciitilde{}radford/fbm.software.html%
}. However, the variance for weights belonging to a given group(either
input-to-hidden weights($u_{ij}$), hidden -biases($a_{j}$), hidden-to-output
weights($v_{j}$) or output-biases($b$)) is chosen to be the same:
$\sigma_{u}^{2}$, $\sigma_{a}^{2}$, $\sigma_{v}^{2}$, $\sigma_{b}^{2}$,
respectively. However, since we don't know, a priori, what these variances
should be, their values are allowed to vary over a large range, while
favoring small variances. This is done by assigning each variance
a gamma prior
\par\end{flushleft}

\begin{center}
\begin{equation}
p\left(z\right)=\left(\frac{\alpha}{\mu}\right)^{\alpha}\frac{z^{\alpha-1}e^{-z\frac{\alpha}{\mu}}}{\Gamma\left(\alpha\right)}\end{equation}

\par\end{center}

\begin{flushleft}
where $z=\sigma^{-2}$, and with the mean $\mu$ and shape parameter
$\alpha$ set to some fixed plausible values. The gamma prior is referred
to as a hyperprior and the parameter of the hyperprior is called a
hyperparameter.
\par\end{flushleft}

\begin{flushleft}
Then, the posterior density, $p\left(\bar{\theta}\mid x,t\right)$,
is gotten according to Eqs. (2),(4) and the prior of Gaussian distribution.
Given an event with data $x'$, an estimate of the target value is
given by the weighted average
\par\end{flushleft}

\begin{center}
\begin{equation}
\bar{y}\left(x'|x,t\right)=\int y\left(x',\bar{\theta}\right)p\left(\bar{\theta}\mid x,t\right)d\bar{\theta}\end{equation}

\par\end{center}

\begin{flushleft}
Currently, the only way to perform the high dimensional integral in
Eq. (6) is to sample the density $p\left(\bar{\theta}\mid x,t\right)$
with the Markov Chain Monte Carlo (MCMC) method\cite{key-1,key-4,key-5,key-6}.
In the MCMC method, one steps through the $\bar{\theta}$ parameter
space in such a way that points are visited with a probability proportional
to the posterior density, $p\left(\bar{\theta}\mid x,t\right)$. Points
where $p\left(\bar{\theta}\mid x,t\right)$ is large will be visited
more often than points where $p\left(\bar{\theta}\mid x,t\right)$
is small.
\par\end{flushleft}

\begin{flushleft}
Eq. (6) approximates the integral using the average
\par\end{flushleft}

\begin{center}
\begin{equation}
\bar{y}\left(x'\mid x,t\right)\approx\frac{1}{L}\sum_{i=1}^{L}y\left(x',\bar{\theta_{i}}\right)\end{equation}

\par\end{center}

\begin{flushleft}
where $L$ is the number of points $\bar{\theta}$ sampled from $p\left(\bar{\theta}\mid x,t\right)$.
Each point $\bar{\theta}$ corresponds to a different neural network
with the same structure. So the average is an average over neural
networks, and is closer to the real value of $\bar{y}\left(x'\mid x,t\right)$,
when $L$ is sufficiently large.
\par\end{flushleft}

\section{Toy Detector and Simulation}

\subsection{Toy Detector}

In the paper, a toy detector is designed to simulate central detectors
in the reactor neutrino experiments, such as Daya Bay experiment\cite{key-7}
and Double Chooz experiment\cite{key-8}, with CERN GEANT4 package\cite{key-9}.
The toy detector consists of three regions, and they are the Gd-doped
liquid scintllator(Gd-LS from now on), the normal liquid scintillator(LS
from now on) and the oil buffer, respectively. The toy detector of
cylindrical shape like the detector modules of Daya Bay experiment
and Double Chooz experiment is designed in the paper. The diameter
of the Gd-LS region is 2.4 meter, and its height is 2.6 meter. The
thickness of the LS region is 0.35 meter, and the thickness of the
oil part is 0.40 meter. In the paper, the Gd-LS and LS are the same
as the scintillator adopted by the proposal of the CHOOZ experiment\cite{key-10}.
The 8-inch photomultiplier tubes (PMT from now on) are mounted on
the inside the oil region of the detector. A total of 366 PMTs are
arranged in 8 rings of 30 PMTs on the lateral surface of the oil region,
and in 5 rings of 24, 18, 12, 6, 3 PMTs on the top and bottom caps.

\subsection{Monte-Carlo Simulation of Toy Detector}

The response of the electron events deposited in the toy detector
is simulated with GEANT4. Although the physical properties of the
scintillator and the oil (their optical attenuation length, refractive
index and so on) are wave-length dependent, only averages\cite{key-10}
(such as the optical attenuation length of Gd-LS with a uniform value
is 8 meter and the one of LS is 20 meter) are used in the detector
simulation. The program couldn't simulate the real detector response,
but this won't affect the result of the comparison between BNN and
MLD. The program allows us to simulate the detector response for the
electron events of the different energy and vertex. In the paper,
10000 electron events regarded as the training sample are uniformly
generated throughout Gd-LS region and their energy is also uniformly
generated from 1 MeV to 13 MeV. 3000 electron events regarded as the
1 MeV test sample are generated uniformly throughout Gd-LS region.
The test samples from 2 MeV to 8 MeV are generated in the same way,
respectively.

\section{Event Reconstruction}

The task of the event reconstruction in the reactor neutrino experiments
is to reconstruct the energy and the vertex of a signal. The maximum
likelihood method is a standard algorithm of the event reconstruction
in the reactor neutrino experiments. The likelihood is defined as
the joint Poisson probability of observing a measured distribution
of photoelectrons over the all PMTs for given ($E,\overrightarrow{x}$)
coordinates in the detector. The Ref.\cite{key-11} for the work of
the CHOOZ experiment shows the method of the reconstruction in detail.
The algorithm of BNN is also applied to event reconstruction, and
its result is compare with the one of MLD.

\subsection{Event Reconstruction with MLD}

In the paper, the event reconstruction with the MLD are performed
in the similar way with the CHOOZ experiment\cite{key-11}, but the
detector is different from the detector of the CHOOZ experiment, so
compared to Ref.\cite{key-11}, there are some different points in
the paper:

(1) The detector in the paper consists of three regions, so the path
length from a signal vertex to the PMTs consist of three parts, and
they are the path length in Gd-LS region, the one in LS region, and
the one in oil region, respectively.

(2) Considered that not all PMTs in the detector can receive photoelectrons
when a electron is deposited in the detector, the $\chi^{2}$ equation
is modified in the paper and different from the one in the CHOOZ experiment,
that is, $\chi^{2}=\sum_{N_{j}=0}\bar{N_{j}}+\sum_{N_{j}\neq0}(\bar{N}_{j}-N_{j}+N_{j}log(\frac{N_{j}}{\bar{N_{j}}}))$,
where $N_{j}$ is the number of photoelectrons received by the j-th
PMT and $\bar{N_{j}}$ is the expected one for the j-th PMT\cite{key-11}.

(3) $c_{E}\times N_{total}$ and the coordinates of the charge center
of gravity for the all visible photoelectrons from a signal are
regarded as the starting values for the fit
parameters($E,\overrightarrow{x}$), where $N_{total}$ is the total
numbers of the visible photoelectrons from a signal and $c_{E}$ is
the proportionality constant of the energy $E$, that is,
$E=c_{E}\times N_{total}$. $c_{E}$ is obtained through fitting
$N_{total}$'s of the 1 MeV electron events, and is
$\frac{1}{235/MeV}$ in the paper.

\begin{flushleft}
($E,\overrightarrow{x}$) of the all electron events, including the
test sample and the training sample, are reconstructed using MLD.
\par\end{flushleft}

\subsection{Event Reconstruction with BNN}

In the paper, the Cartesian coordinates ($x,y,z$) of the all events,
including the test sample and the training sample, are transformed
to their cylindrical coordinates ($r,\theta,z$). The ($E,r,\theta,z$)
are used as inputs to the BNN, which have the input layer of 4 inputs,
the single hidden layer of 8 nodes and the output layer of a output
which is $E$, or $x,y,z$, respectively. The $E$ and $x,y,z$ of
the test samples are predicted using the BNN, respectively. A Markov
chain of neural networks is generated using the Bayesian neural networks
package of Radford Neal, with the training sample, in the process
of the event reconstruction. One thousand iterations, of twenty MCMC
steps each, are used in the paper. The neural network parameters are
stored after each iteration, since the correlation between adjacent
steps is very high. That is, the points in neural network parameter
space are saved to lessen the correlation after twenty steps. It is
also necessary to discard the initial part of the Markov chain because
the correlation between the initial point of the chain and the points
of the part is very high. The initial three hundred iterations are
discarded in the paper.

\section{Conclusion}

\begin{flushleft}
Fig.1, Fig.2 and Fig.3 illustrate the results of the event
reconstruction with BNN and MLD. Fig.1 shows that the errors of the
vertex of 1 MeV and 8 MeV electrons reconstructed by the BNN are
consistent with the ones by MLD, that is, they are not obviously
different. Fig.2 shows that the energy uncertainty for 1 MeV
electrons with BNN decreases by 95.0\% in comparison with the one
using MLD. And the uncertainty in the case of the 8 MeV events
decreases by 76.3\%. Fig.3 shows the energy resolutions using BNN
are more significantly improved in comparison with the one using MLD
while increasing energy. Meanwhile, the relative errors of the
energy resolutions are about 2.0\%, and are from fit errors (about
1.5\%) and statistical errors (about 1.3\%). So the difference
between results of BNN and MLD is not significant in the case of 1
MeV events in consideration of the effect of statistical
fluctuations. But the contribution to the difference is mainly from
the superiority of BNN for the events from 2 MeV to 8 MeV. Thus it
can be seen that the energy resolutions using BNN are significantly
improved for the high energy events in comparison with the one using
MLD. Therefore, BNN can be well applied to the energy reconstruction
in the reactor neutrino experiments, and the better energy
resolution is obtained by BNN.
\par\end{flushleft}

\begin{flushleft}
Although the discussion in the paper are only for the reactor neutrino
experiments, it is expected that the algorithm of BNN can also be
applied to the event reconstruction of the other experiments and will
find wide application in the experiments of high energy physics.
\par\end{flushleft}

\section{Acknowledgements }

This work is supported by the National Natural Science Foundation
of China(NSFC) under the contract No. 10605014.

\newpage{}

\begin{figure}
\includegraphics[width=16cm,height=16cm]{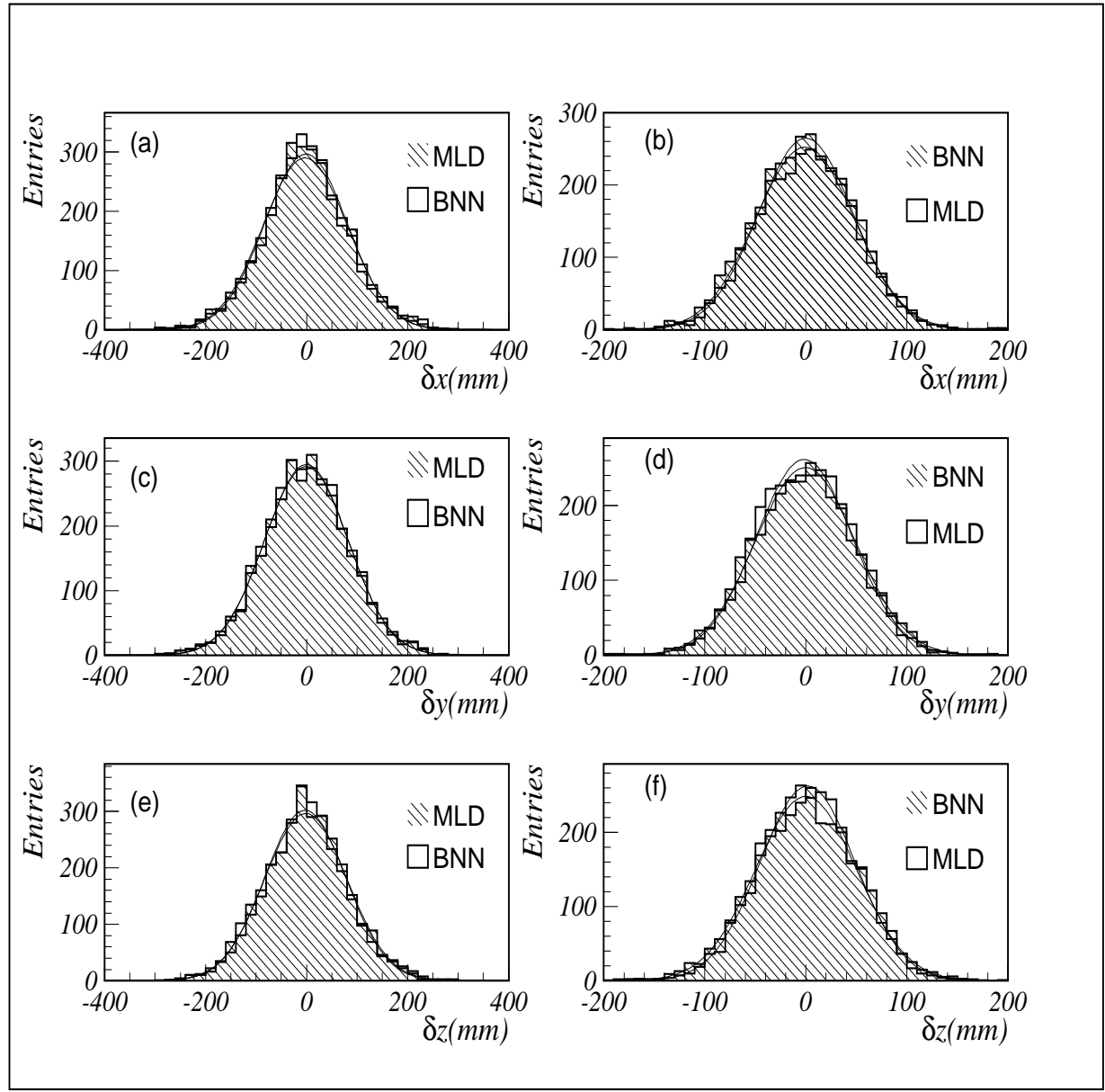}

\caption{$\delta x,\delta y,\delta z$ is the difference between the coordinates
of the reconstructed position and the generated ones, respectively.
The event position is reconstruted using BNN and MLD, respectively.
(a)(c)(e) illustrate the difference distribution of the 1 MeV electrons,
and (b)(d)(f) illustrate the one of 8 MeV electrons.}
\end{figure}

\newpage{}

\begin{figure}
\includegraphics[width=16cm,height=16cm]{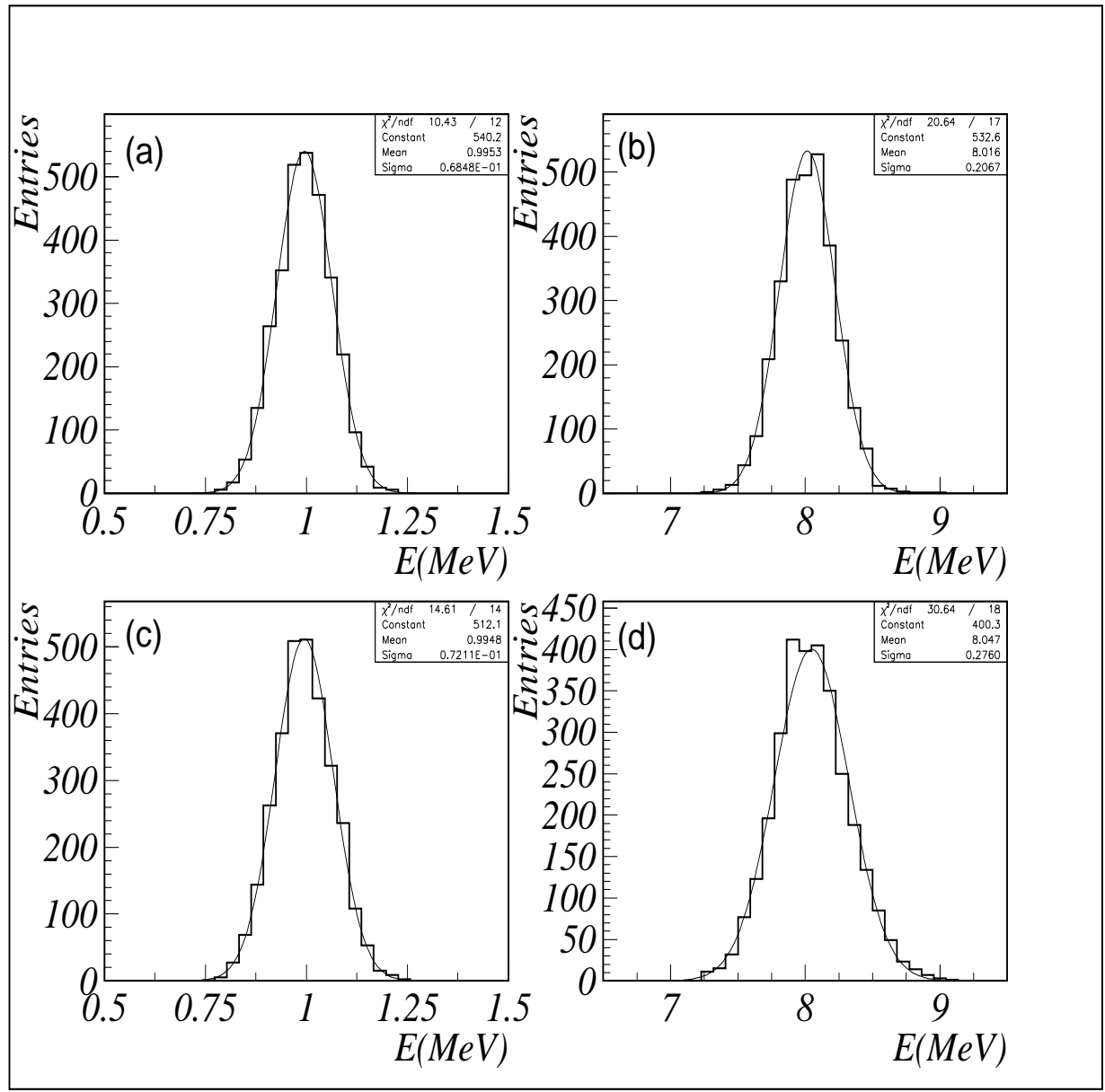}

\caption{The energies of 1 MeV and 8 MeV electrons are reconstructed using
BNN and MLD, respectively. (a), (b) illustrate the distribution of
the energy reconstructed by BNN for 1 MeV and 8 MeV electrons, respectively.
(c), (d) illustrate the distribution of the energy reconstructed by
MLD for 1 MeV and 8 MeV electrons, respectively.}
\end{figure}

\newpage{} %
\begin{figure}
\includegraphics[width=16cm,height=16cm]{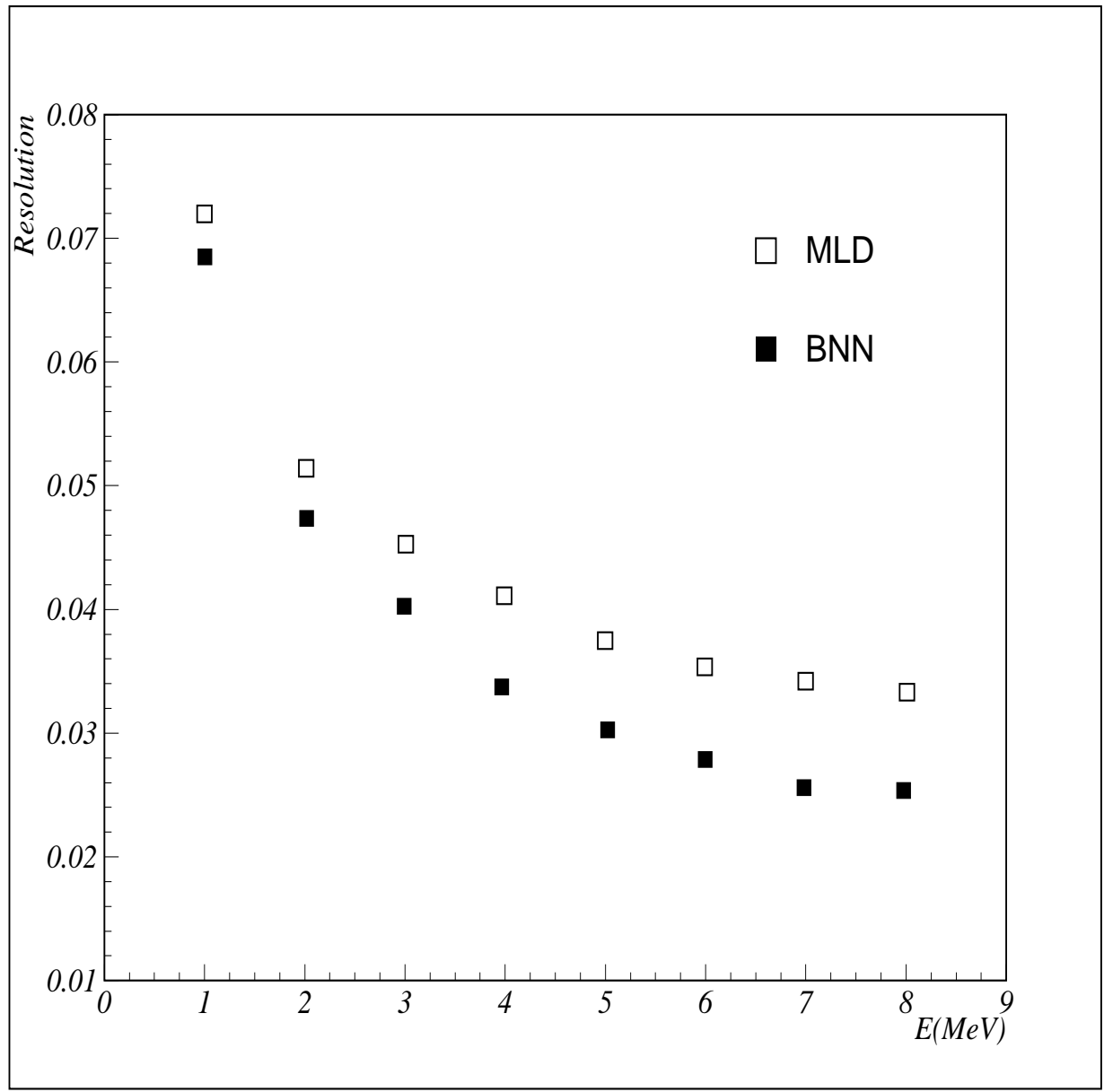}

\caption{The energy resolution of the test sample from 1 MeV to 8 MeV are
shown in the figure. The white squares denote the resolutions using
MLD, and the black squares denote the one using BNN.}
\end{figure}

\end{document}